\documentclass[aps,pra,reprint,groupaddress]{revtex4-1}
\usepackage[utf8]{inputenc}
\usepackage{bm}
\usepackage{enumerate}
\usepackage{graphicx}
\usepackage{amsmath}
\usepackage{amssymb}
\usepackage{cancel}
\usepackage{xcolor}
\usepackage{verbatim}
\usepackage{hyperref}
\usepackage{lipsum}
\usepackage{bbold}
\usepackage{comment}
\usepackage{scalerel}
\hypersetup{colorlinks=true,
linkcolor=blue,
citecolor=blue,
urlcolor=blue,
filecolor=blue
}

\def\ket#1{|{#1}\rangle}

\newcommand{\bs}{\boldsymbol}
\allowdisplaybreaks

\begin{abstract}
Qubits encoded in the spin state of heavy holes confined in Si- and Ge-based semiconductor quantum dots are currently leading the efforts toward spin-based quantum information processing.
The virtual absence of spinful nuclei in purified samples yields long qubit coherence times and the intricate coupling between spin and momentum in the valence band can provide very fast spin--orbit-based qubit control, e.g., via electrically induced modulations of the heavy-hole $g$-tensor.
A thorough understanding of all aspects of the interplay between spin--orbit coupling, the confining potentials, and applied magnetic fields is thus quintessential for the development of the optimal hole-spin-based qubit platform.
Here we theoretically investigate the manifestation of the effective $g$-tensor and effective mass of heavy holes in two-dimensional hole gases as well as in lateral quantum dots.
We include the effects of the anisotropy of the effective Luttinger Hamiltonian (particularly relevant for Si-based systems) and we focus on the detailed role of the orientation of the transverse confining potential.
We derive most general analytic expressions for the anisotropic $g$-tensor and we present a general and straightforward way to calculate corrections to this $g$-tensor for localized holes due to various types of spin--orbit interaction, exemplifying the approach by including a simple linear Rashba-like term.
Our results thus contribute to the understanding needed to find optimal points in parameter space for hole-spin qubits, where confinement is effective and spin--orbit-mediated electric control over the spin states is efficient.
\end{abstract}

\begin{document}

\title{Anisotropic \textit{g}-tensors in hole quantum dots:\\ The role of the transverse confinement direction}

\author{J{\o}rgen Holme Qvist}
\affiliation{Center for Quantum Spintronics, Department of Physics, Norwegian University of Science and Technology, NO-7491 Trondheim, Norway}

\author{Jeroen Danon}
\affiliation{Center for Quantum Spintronics, Department of Physics, Norwegian University of Science and Technology, NO-7491 Trondheim, Norway}

\date{\today}

\maketitle

\section{Introduction}
Electron-spin-based qubits hosted in gate-defined semiconductor quantum dot structures have long been a promising candidate for easily scalable quantum information processors~\cite{Loss1998,Hanson2007,Chatterjee2021}.
Although GaAs-based devices have propelled the field forward for more than a decade, yielding many encouraging features such as full electric control and fast operation times~\cite{DiVincenzo2000,Russ2017,Laird2010,Gaudreau2011,Medford2013,Medford2013a}, their coherence times are intrinsically limited due to the coupling between the electron spins and the nuclear spin bath of the host material~\cite{Khaetskii2002,Hung2014,Peterfalvi2017}.
A potential solution to this problem is to host the qubits in group-IV materials, such as Si or Ge, which can be made almost nuclear-spin-free by isotopic purification~\cite{Veldhorst2014,Muhonen2014,Eng2015,Yoneda2018,Andrews2019}.
However, this approach comes with the complication of an extra valley degree of freedom for confined electrons, which is hard to control and provides an extra channel for leakage and dephasing~\cite{Zwanenburg2013,Culcer2010a}.

Lately there has been dramatic progress with Si- and Ge-based spin qubits that use instead of electron spin the spin of valence-band holes~\cite{Maurand2016,Liles2018,Watzinger2018,Vukusic2018,Crippa2019,Scappucci2020,Jirovec2021,Lawrie2021,Hendrickx2021}.
These holes provide a similar protection against magnetic noise as the electrons, due to the virtual absence of nuclear spins in purified samples, but they do not have the complicating valley degree of freedom.
However, since the orbitals that constitute the valence band are of $p$-type~\cite{winkler2003}, the corresponding states have a total six-fold angular momentum degree of freedom, possibly leading to highly anisotropic dynamics.
Compared to the valley mixing of the electronic states, however, these dynamics are relatively predictable, and the built-in mixing of orbital and spin degrees of freedom can yield strong effective spin--orbit coupling that allows for fast qubit operation~\cite{ares2013,Kloeffel2011,Kloeffel2018,Bosco2021,Bosco2021a,Froning2021a,Wang2021}.
Moreover, the $p$-type orbital nature of the valence band has the additional advantage of weaker effective hyperfine coupling to any residual spinful nuclei, due to the wave function having a node at the atomic site \cite{Fischer2008}.

Recent experiments on two-dimensional hole quantum wells and quantum dots have indeed shown wildly varying and anisotropic effective hole masses \cite{Chiu2011,Hardy2019,Lodari2019} and $g$-factors \cite{Koduvayur2008,Voisin2016,watzinger2016,Brauns2016,Bogan2017,Lu2017,Crippa2018,Gradl2018,Vries2018,Sammak2019,hofmann2019ArXiV,Miller2021}, depending on choice of material, hole densities, and on the details of the confinement.
%
In this paper we theoretically investigate these anisotropic properties of confined holes in detail, with a focus on the role of the precise orientation of the confinement potentials with respect to the crystal orientation.
We will pay special attention to the case of Si, which has particularly strong anisotropic properties as compared to most other common materials, such as Ge, GaAs and InAs~\cite{winkler2003}.

We assume a semiconductor heterostructure containing a thin layer to which a two-dimensional hole gas (2DHG) is confined.
Further in-plane confinement into quantum dots can then be realized using electrostatic top gates.
We do not restrict our analysis to confinement planes along the common crystal growth directions, but investigate the more general case where the 2DHG can be oriented along any arbitrary direction.
Although from a fabricational point of view it is maybe not straightforward to realize arbitrary confinement directions (or directions that are incommensurate with the primitive lattice vectors), exploring the fully general case will allow to identify orientations with optimally tuned parameters for spin-qubit implementations in different materials, such as, e.g., minimal effective masses, maximal in-plane $g$-factors, and maximal electrical tunability of the $g$-tensor.
Apart from revealing analytical insights in the relation between the orientation of the 2DHG and the most important effective parameters of the resulting quantum dots, our results could thus also serve as inspiration for exploring possibilities to create confinement planes in less common crystallographic directions.

In Sec.~\ref{sec:Luttinger_Hamiltonain} we present the effective $4\times 4$ Luttinger Hamiltonian we use to describe the hole dynamics in the top part of the valence band.
We then write the Hamiltonian as a function of two Euler (rotation) angles such that the $z$-direction can be made to point in any desired direction.

In Sec.~\ref{sec:2DHG} we add a strong confinement potential along the (arbitrary) $z$-direction, assuming that the corresponding confinement energy scale dominates all other relevant scales in the system.
The terms in the Luttinger Hamiltonian that break spherical symmetry allow the confinement to mix states with different angular momentum along the out-of-plane direction, thereby introducing anisotropy in the effective hole parameters.
Under the assumption of strong confinement we thus diagonalize the dominating part of the Hamiltonian and extract analytic expressions for the in-plane effective hole masses.
Then we add a Zeeman Hamiltonian describing the coupling of the hole spins to an applied magnetic field.
We transform this Hamiltonian to the same ``eigenbasis'' defined by the transverse confinement and extract analytic expressions for the full anisotropic $g$-tensor for the lowest two hole spin states, as a function of the two Euler angles.
We map out the full orientation-dependence of the $g$-tensor for the case of Si, showing a great variation of magnitude and sign in its components.


Finally, in Sec.~\ref{sec:Quantum_dots}, we add in-plane confinement into quantum dots, assuming the corresponding orbital energy scale to be much smaller than the out-of-plane orbital energy.
Although we assume a circularly symmetric confining potential, the anisotropic effective hole mass makes the confinement effectively elliptic.
We add the effect of the out-of-plane component of the applied magnetic field and use a diagonalized version of a single quantum-dot Hamiltonian in terms of bosonic ladder operators.
Expressing the hole momentum operators in terms of the same ladder operators allows for a straightforward and versatile perturbative evaluation of the effect of spin--orbit interaction (SOI) on the dynamics of the confined holes.
We exemplify this approach by including a simple linear Rashba-like SOI that can result from the out-of-plane confinement, and we derive analytic expressions for the resulting corrections to the $g$-tensor for the confined holes.
As we point out below, including other types of SOI that might dominate depending on choice of material and details of confinement is simple in our approach, and our results can straightforwardly be used to produce analytic expressions for the $g$-tensor corrections due to any desired type of SOI.
The SOI-induced corrections to the $g$-tensor of localized holes can used for fast spin manipulation through electrical $g$-tensor modulation~\cite{Voisin2016,Crippa2018,Kato2003,Venitucci2018}, and developing a thorough understanding of the detailed interplay of SOI, confinement, and applied magnetic fields is thus crucial~\cite{Terrazos2021,Michal2021,Adelsberger2021}.

\section{The Hamiltonian}
\label{sec:Luttinger_Hamiltonain}
In semiconductors with diamond or zinc-blende structure the states in the valence band are comprised of atomic orbitals with angular momentum $l=1$ and spin $s = \frac{1}{2}$.
The band thus has a six-fold degree of freedom that can be classified in terms of total angular momentum $j = l+s$.
Spin--orbit interaction splits off the two states with $j = \frac{1}{2}$ from the other four by an energy of the order $\sim 100$~meV, and for the low-energy dynamics one can thus focus on the four $j=\frac{3}{2}$ states.
Using $\bs k\cdot\bs p$ theory, one can derive an effective $4\times 4$ Hamiltonian for this subspace, which reads in the cubic approximation
\begin{align}
    H_\text{L}= {} & {} \frac{p^2}{2m_0}\left(\gamma_1+\frac{5}{2}\gamma_2\right) - \frac{\gamma_2}{m_0}\left(p_x^2J_x^2 + \text{c.p.} \right)
    \nonumber\\&
    - \frac{2\gamma_3}{m_0}\big( \{p_x,p_y\}\left\{J_x,J_y\right\} + \text{c.p.} \big),
    \label{eq:bulk_Luttinger}
\end{align}
where $\{A,B\}=\frac{1}{2}\left(AB+BA\right)$, $m_0$ is the electron rest mass, $p_i$ are the momentum operators, with $i \in \{x,y,x\}$, $J_{i}$ are the three spin-$\frac{3}{2}$ matrices, and c.p. denotes cyclic permutation. Furthermore, the dimensionless constants $\gamma_{1,2,3}$ are the three so-called Luttinger parameters, and are given in Tab.~\ref{tab:Luttinger_parameters} for Si, Ge, GaAs and InAs. 

Although we will mainly focus on the dynamics governed by the Luttinger Hamiltonian (\ref{eq:bulk_Luttinger}), the effect of strain could easily be added by including the so-called Bir-Pikus Hamiltonian \cite{Sun2009},
\begin{align}
    H_\text{BP} = {} & {} \left(-a+\frac{5}{4}b\right)\left(\epsilon_{xx}+\text{c.p.}\right) - b\left(\epsilon_{xx}J_x^2 + \text{c.p.}\right) 
    \nonumber\\&
    - \frac{2d}{\sqrt{3}}\big( \epsilon_{xy}\left\{J_x,J_y\right\} + \text{c.p.} \big),
    \label{eq:bir-pikus}
\end{align}
where $\bar\epsilon$ is the strain tensor, $a$ is the Bir-Pikus hydrostatic deformation potential, and $b$ and $d$ are two Bir-Pikus shear deformation potentials \cite{Sun2009}.
This Hamiltonian has the same structure as the Luttinger Hamiltonian \eqref{eq:bulk_Luttinger}, which in principle allows for a straightforward inclusion of strain into the results we will report below.

\begin{table}[b]
    \centering
    \caption{Luttinger parameters $\gamma_{1,2,3}$ and bare effective $g$-factors $\kappa$ and $q$ in Si, Ge, GaAs, and InAs~\cite{winkler2003}.}\label{tab:Luttinger_parameters}
    \addtolength{\tabcolsep}{2pt}
    \begin{tabular}{l|cccc}
         & Si & Ge & GaAs & InAs \\ \hline
        $\gamma_1$ & \phantom{$-$}4.285 & 13.38 & 6.85 & 20.40 \\
        $\gamma_2$ & \phantom{$-$}0.339 & \phantom{2}4.24 & 2.10 & \phantom{2}8.30 \\
        $\gamma_3$ & \phantom{$-$}1.446 & \phantom{2}5.69 & 2.90 & \phantom{2}9.10 \\
        $\kappa$ & $-0.42$\phantom{2} & \phantom{2}3.41 & 1.20 & \phantom{2}7.60 \\
        $q$ & \phantom{$-$}0.01\phantom{2} & \phantom{2}0.06 & 0.01 & \phantom{2}0.39
    \end{tabular}
	\addtolength{\tabcolsep}{-2pt}
\end{table}

In both Hamiltonians \eqref{eq:bulk_Luttinger} and \eqref{eq:bir-pikus} it is assumed that the coordinate system $\{x,y,z\}$ is aligned with the main crystallographic axes.
This is important since the two last terms are not spherically symmetric, i.e., the structure of these two terms depends on the choice of coordinate system.
In many common semiconductors such as GaAs, Ge, and InAs the difference $|\gamma_2 - \gamma_3|$ is smaller than both $\gamma_2 + \gamma_3$ and $\gamma_1 + \frac{5}{2}\gamma_2$ (see Tab.~\ref{tab:Luttinger_parameters}), which makes setting $\gamma_2$ equal to $\gamma_3$ a reasonable approximation.
In that case, the Hamiltonian becomes spherically symmetric and no longer depends on the orientation of the coordinate system with respect to the crystal structure.
However, since we specifically want to include Si in our consideration, for which the spherical approximation is not particularly good, we will not set $\gamma_2$ and $\gamma_3$ equal, and will take the actual crystal orientation into account.

A 2DHG is created by applying strong confinement along one direction.
To find an effective in-plane two-dimensional Hamiltonian for the 2DHG we need to integrate out the coordinate along the direction of confinement, which we will call $z$.
If $z$ does not point along one of the main crystallographic axes, we first need to rotate the Hamiltonian to the correct coordinate system.
This can be done as follows:
(i)~The original Hamiltonian is separated in a spherically symmetric part, which is invariant under rotations, and a cubic part that is comprised of the $0$ and $\pm 4$ components of the rank-4 part of the tensor product of the two irreducible rank-2 tensors that can be formed from the elements $K_{ij} = \frac{3}{2}(p_ip_j+p_jp_i) -\delta_{ij}p^2$ and $L_{ij} = \frac{3}{2}(J_iJ_j + J_jJ_i) - \delta_{ij} J^2$~\cite{Fishman1995}.
(ii) The cubic contribution can be rotated to the new coordinate system by applying the rotation matrix for $j=4$ angular-momentum eigenfunctions ${\bf D}^{(4)}(\alpha,\beta,\gamma)$ to the components of the rank-4 tensor mentioned above, where $\{\alpha, \beta,\gamma\}$ are the Euler angles of the rotation~\cite{tinkham1964}.
In this work we will explore the full range of possible confinement planes and thus not restrict ourselves to the common crystal growth directions such as $[nnm]$.

\begin{figure}[tb]
	\centering
	\includegraphics[width=.6\linewidth]{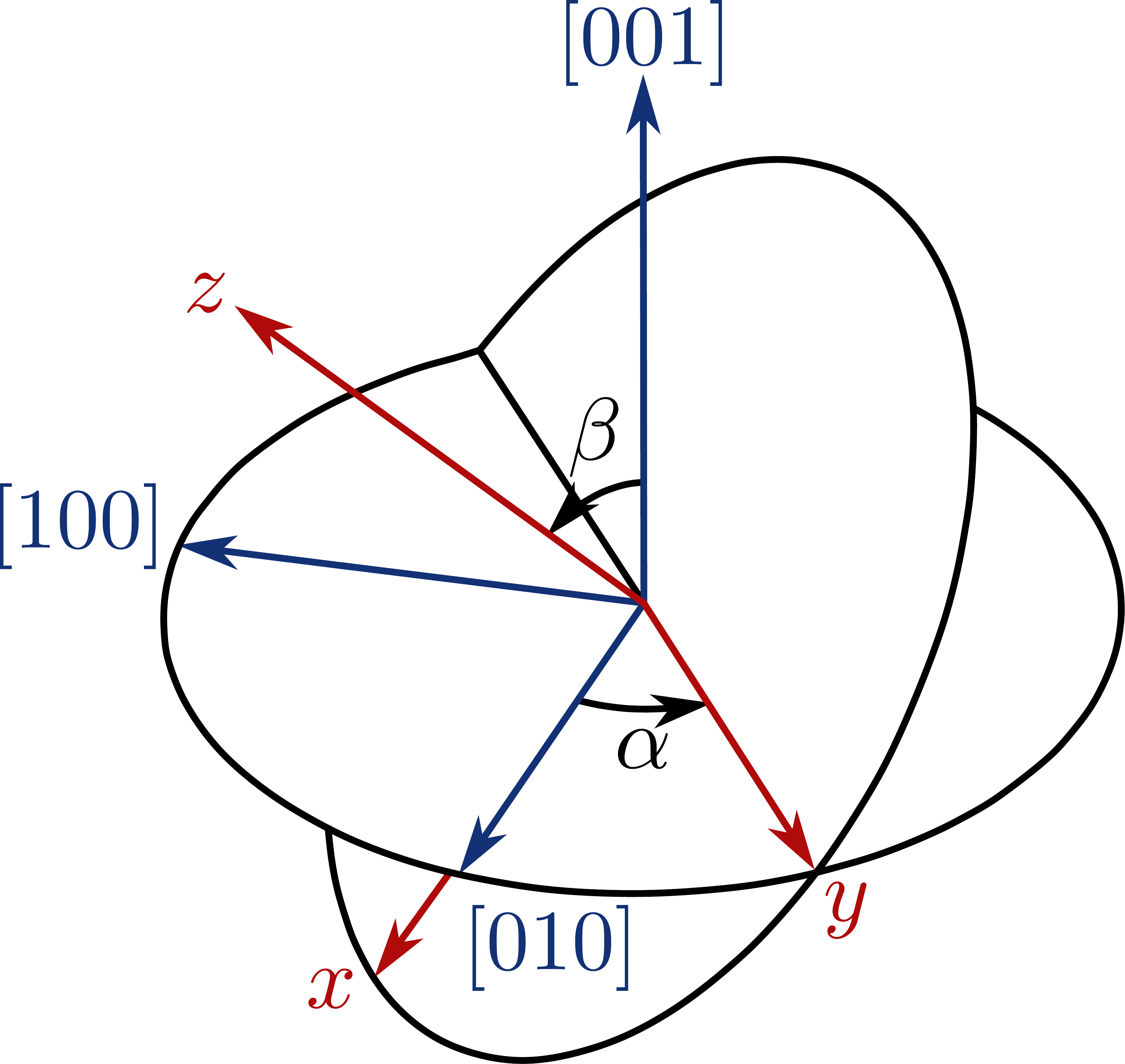} 
	\caption{Illustration of the rotation using the two Euler angles $\alpha$ and $\beta$. The crystallographic axes are shown in blue, the rotated coordinate system $(x,y,z)$ is shown in red.}
	\label{fig:Euler_angles}
\end{figure}

Since any plane of confinement can be defined by two angles only, we fix $\gamma=0$ to simplify our analytic expressions.
The new coordinate system then results from a rotation over $\alpha$ about [001] followed by a rotation over $\beta$ about the new $y$-axis, as illustrated in Fig.~\ref{fig:Euler_angles}.
In that way the $[nnm]$-directions, as investigated in \cite{winkler2003,Fishman1995}, can be obtained by simply setting $\alpha=\pi/4$.
Most experiments use samples grown along the [001] and [110] directions, with the confinement created along the growth direction, and when presenting explicit results we will thus consider these highly used confinement directions.
However, since we can obtain results for any general direction of confinement, it is straightforward to also explore less common directions, which could result in a 2DHG with more interesting or useful properties.

The resulting rotated Hamiltonian can always be written in the following form,
\begin{equation}
H(\alpha,\beta) = \left(\begin{array}{cccc}
P-Q & -S & R & 0 \\
-S^\dagger & P+Q & 0 & R \\
R^\dagger & 0 & P+Q & S \\
0 & R^\dagger & S^\dagger & P-Q
\end{array}\right),
\label{eq:Ham_mmn}
\end{equation}
in the basis of the eigenstates $\{\ket{\frac{3}{2}},\ket{\frac{1}{2}},\ket{-\frac{1}{2}},\ket{-\frac{3}{2}}\}$ of $J_z$ with its quantization axis along the new $z$-direction.
The matrix elements $P$, $Q$, $R$ and $S$ can be expressed in terms of dimensionless symmetric tensors $M_{ij}$,
\begin{equation}
M = \frac{1}{2m_0}\sum_{i,j}M_{ij}\{ p_i, p_j \},
\end{equation}
where $M\in\{P,Q,R,S\}$ and $i,j\in\{x,y,z\}$.
The diagonal element $P$ is invariant under rotations and follows from $P_{ij} = \delta_{ij}\gamma_1$;
the other elements are more involved and explicit expressions for their $M_{ij}$ as a function of $\alpha$ and $\beta$ are given in App.~\ref{app:Luttinger_elements}.
The Bir-Pikus contribution to the Hamiltonian can easily be included in this notation, by adding a similar contribution $M \to M+\sum_{i,j}M^{\rm BP}_{ij}\epsilon_{ij}$, where the elements $M^{\rm BP}_{ij}$ can be obtained from $M_{ij}$ by the substitution $\{\gamma_1,\gamma_2,\gamma_3\} \to \{-a,\frac{1}{2}b, \frac{1}{2\sqrt 3}d\}$.

\section{The 2-dimensional hole gas}
\label{sec:2DHG}

In this section we investigate the dynamics of the holes in a 2DHG, and calculate their effective masses and $g$-tensor.
The in-plane Hamiltonian for the confined holes is obtained by integrating out the coordinate along the direction of confinement, which we labeled $z$.
Assuming no strain and an infinite-well type of confinement for simplicity, one finds that all terms in $H$ that are linear in $p_z$ vanish, also in the presence of a finite magnetic field~\cite{winkler2003},
and the terms quadratic in $p_z$ integrate out to contributions $M_{zz}u_{z}$, where the confinement energy scale $u_{z} = \langle p_z^2 \rangle/2m_0$ will be assumed much larger than the in-plane kinetic energy of the holes.

The next step is to diagonalize the part of the Hamiltonian that is proportional to $u_z$, which in general leads to a basis that no longer consists of pure $m_j=\pm\frac{3}{2}$ and $m_j=\pm\frac{1}{2}$ states.
The two resulting pairs of spin-mixed eigenstates are the heavy and light holes (HHs and LHs), where the heavy holes are the ones with the lowest excitation energy.
The light holes are split off by an energy $\Delta_{\rm HL} = 2u_z\sqrt{Q_{zz}^2 + |R_{zz}|^2 + |S_{zz}|^2}$, but can become mixed with the heavy holes by in-plane confinement or an applied magnetic field.

\subsection{Effective masses}
\label{sec:2DHG_beyond}

\subsubsection{Spherical approximation, $\delta=0$}

Before investigating the anisotropic dynamics of the 2DHG, we briefly repeat the well-known results for the spherical approximation, which follow from setting $\delta\equiv\gamma_3-\gamma_2 \to 0$.
Then, the HH and LH states at the band edge are pure $m_j=\pm\frac{3}{2}$ and $m_j=\pm\frac{1}{2}$ states, and one finds that $S=0$ and
\begin{align}
    Q &= \frac{4\gamma_2+6\gamma_3}{5}u_z -\frac{1}{10m_0}(2\gamma_2+3\gamma_3)(p_x^2+p_y^2),
    \label{eq:Q_spherical}
    \\
    R &= -\frac{\sqrt{3}}{10m_0}(2\gamma_2+3\gamma_3)(p_x-ip_y)^2,
    \label{eq:R_spherical}
\end{align}
so that $\Delta_\text{HL} = \frac{4}{5}\left(2\gamma_2+3\gamma_3\right)u_z$.
We see that the Hamiltonian is indeed spherically symmetric in this limit and irrespective of the crystallographic orientation of the 2DHG the in-plane effective masses read to leading order in $1/u_z$ as $m^\text{H(L)} = m_0/\left[\gamma_1\pm\frac{2}{5}(2\gamma_2+3\gamma_3)\right]$ for the HHs and LHs, respectively.

\subsubsection{Anisotropic Hamiltonian, $\delta\neq0$}

The spherical approximation is good in materials where $\delta/(2\gamma_2+3\gamma_3)$ is very small.
For the case of Si, however, we have $\delta/(2\gamma_2+3\gamma_3) \approx 0.22$ which is not negligible, and we thus need to include the terms proportional to $\delta$ as well.
In general this results in the HHs and LHs becoming mixtures of the $m_j=\pm\frac{3}{2}$ and $m_j=\pm\frac{1}{2}$ states, except for confinement along high-symmetry directions, such as [001] and [111], where $R_{zz}=S_{zz}=0$ and the Hamiltonian becomes isotropic again.

\begin{figure}[tb]
\centering
\includegraphics[width=\linewidth]{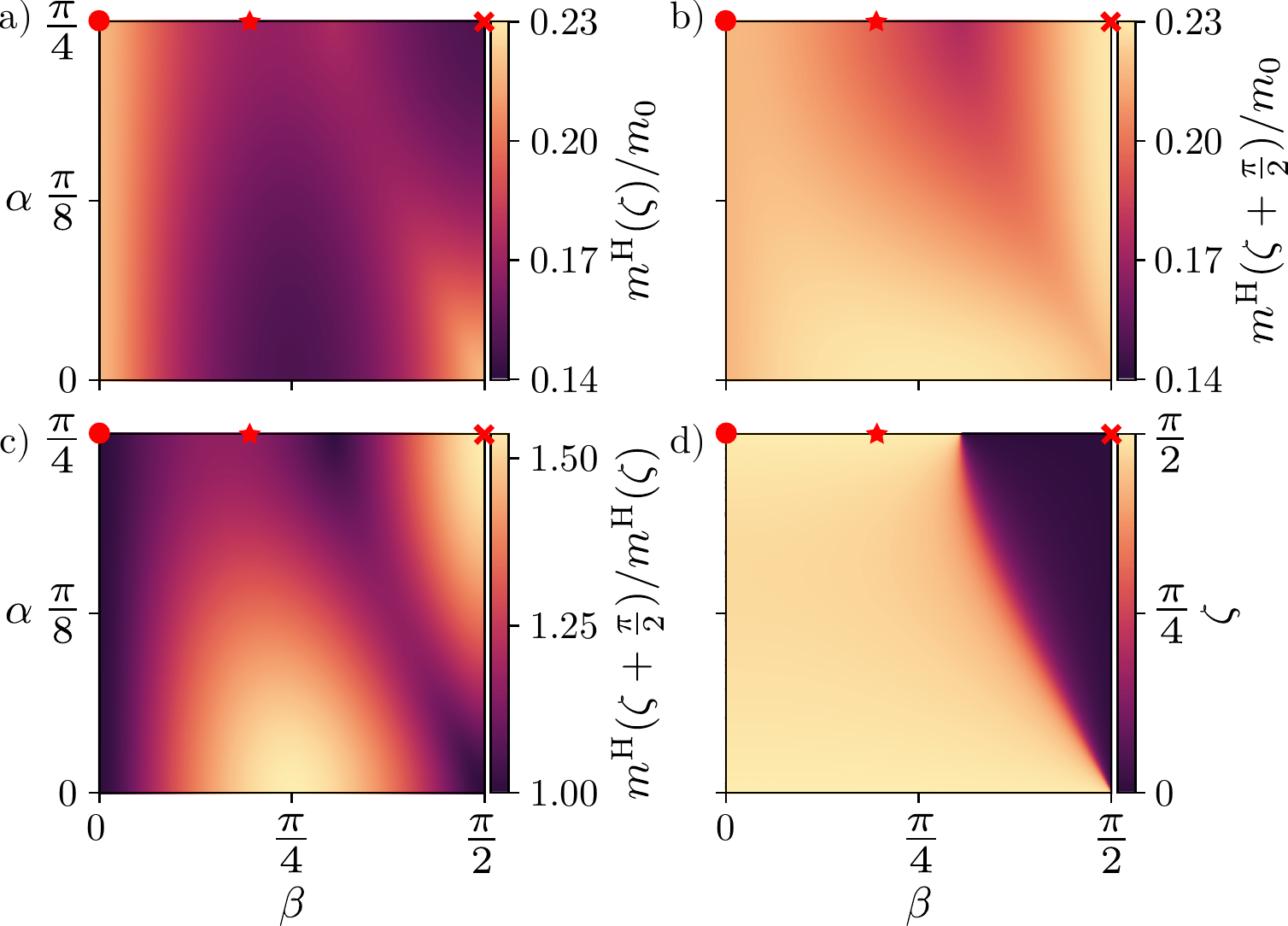} 
\caption{The effective hole masses of Si as predicted by Eq.~\eqref{eq:masses}, plotted against the direction of confinement. (a) and (b) show the smallest and largest effective mass $m^\text{H}(\zeta)$ and $m^\text{H}(\zeta+\pi/2)$, respectively, (c) shows the anisotropy of the effective masses $m^\text{H}(\zeta+\pi/2)/m^\text{H}(\zeta)$, and (d) shows the angle $\zeta$.}
\label{fig:effective_masses}
\end{figure}

Most generally, we find for the in-plane effective masses to leading order in $1/u_z$
\begin{align}
m^{{\rm H},{\rm L}}(\theta) = \frac{2m_0}{2\gamma_1 + a + b \cos(2\theta -2\zeta)},
\label{eq:masses}
\end{align}
where $\theta$ is the angle between the $x$-axis and the direction of motion of the hole.
The parameters $a$ and $b$ are different for the HHs and LHs,
\begin{align*}
a^{{\rm H},{\rm L}} = {} & {} \mp \text{Re}[ {\bs n}_{zz} \cdot ({\bs v}_{xx} + {\bs v}_{yy})^*],\\
b^{{\rm H},{\rm L}} = {} & {} \pm \sqrt{ \text{Re}[ {\bs n}_{zz} \cdot ({\bs v}_{xx} - {\bs v}_{yy})^*]^2 +\text{Re}[ {\bs n}_{zz} \cdot {\bs v}_{xy}^*]^2 },
\end{align*}
where we introduced the vectors ${\bs v}_{\alpha\beta} \equiv \{ Q_{\alpha\beta}, R_{\alpha\beta}, S_{\alpha\beta}\}$ and ${\bs n}_{\alpha\beta} \equiv {\bs v}_{\alpha\beta} / |{\bs v}_{\alpha\beta}|$.
The angle
\begin{equation}
\zeta = \frac{1}{2}\arctan\left(\frac{\text{Re}[ {\bs n}_{zz} \cdot {\bs v}_{xy}^*]}{2 \text{Re}[ {\bs n}_{zz} \cdot ({\bs v}_{yy} - {\bs v}_{xx})^*]}\right)
\label{eq:angle_zeta}
\end{equation}
determines what $\theta$ gives the smallest(largest) effective heavy(light) hole mass, while the largest(smallest) effective mass is always obtained when $\theta$ is an angle $\frac{\pi}{2}$ off from $\zeta$. 
Inserting the expressions given in App.~\ref{app:Luttinger_elements} reveals the explicit dependence of $m^{{\rm H},{\rm L}}(\theta)$ on the Euler angles that were used to rotate the Hamiltonian. 

In Fig.~\ref{fig:effective_masses} we illustrate how the effective HH masses in Si depend on the two Euler angles $\alpha$ and $\beta$. 
Fig.~\ref{fig:effective_masses}(a) and (b) show the magnitudes of the smallest effective mass $m^\text{H}(\zeta)/m_0$ and the largest effective mass $m^\text{H}(\zeta+\frac{\pi}{2})/m_0$, respectively. In (c) we plot the anisotropy of the effective masses $m^\text{H}(\zeta+\frac{\pi}{2})/m^\text{H}(\zeta)$, while (d) shows how the angle $\zeta$ depends on $\alpha$ and $\beta$.

The three symbols in Fig.~\ref{fig:effective_masses} mark three common confinement directions: [001] (circle), [110] (cross) and [112] (star).
For the high-symmetry direction [001] the effective masses are isotropic, as expected.
Inserting zero for $\alpha$ and $\beta$ in Eq.~\eqref{eq:masses} we find $m^\text{H,L}_{[001]}(\theta) = m_0/(\gamma_1\pm\gamma_2)$, which is indeed independent on $\theta$.
The largest anisotropy of the effective masses is obtained when the 2DHG is confined along [110], where $m^\text{H}(\zeta+\frac{\pi}{2})/m^\text{H}(\zeta)>1.5$. Here, the parameter $b$ in Eq.~(\ref{eq:masses}) is at its maximum, making the masses highly dependent on $\theta$,
\begin{equation}
    m^\text{H,L}_{[110]}(\theta) = \frac{2m_0}{2\gamma_1\pm\sqrt{\gamma_2^2+3\gamma_3^2}\pm3\frac{|\gamma_3^2-\gamma_2^2|}{\sqrt{\gamma_2^2+3\gamma_3^2}}\cos(2\theta)}.
\end{equation}
We find a similar expression for the [112] direction, 
\begin{equation}
    m^\text{H,L}_{[112]}(\theta) = \frac{2m_0}{2\gamma_1\pm\sqrt{\gamma_2^2+3\gamma_3^2}\mp\frac{|\gamma_3^2-\gamma_2^2|}{\sqrt{\gamma_2^2+3\gamma_3^2}}\cos(2\theta)}.
\end{equation}
which has less anisotropy and opposite directions where the masses are largest and smallest, as compared to [110].


\subsection{Heavy-hole Zeeman effect}
\label{sec:g-tensor}

We will now add a magnetic field and consider its coupling to the angular momentum of the HHs in a 2DHG through the Zeeman effect.
The Hamiltonian describing this coupling for the four $j=\frac{3}{2}$ states in the upper valence band reads as \cite{Winkler2000,winkler2003}
\begin{equation}
    H_{\rm Z}= 2\kappa \bs B\cdot\bs J + 2q\bs B\cdot\mathbf{\mathcal{J}},
    \label{eq:zeeman}
\end{equation}
where $\kappa$ is the effective $g$-factor of the isotropic coupling, $\bs B$ is the applied magnetic field, $q$ sets the strength of the anisotropic coupling, $\mathcal{J}=\left\{J_x^3,J_y^3,J_z^3\right\}$, and we use units where the Bohr magneton $\mu_{\rm B}=1$.
Since $\kappa$ is usually two orders of magnitude larger than $q$ (see Tab.~\ref{tab:Luttinger_parameters}) we will neglect the anisotropic contribution to $H_{\rm Z}$.
The goal of this section is to derive an effective $g$-tensor $\bar g$ for the HH subspace, such that the linear Zeeman Hamiltonian (\ref{eq:zeeman}) for the HHs can be written as
\begin{align}
H_{\rm Z}^\text{H} = \frac{1}{2} \bs \sigma \cdot\bar g \cdot\bs B,
\label{eq:Zeeman_HH}
\end{align}
where $\bs\sigma = \{\sigma_x, \sigma_y, \sigma_z \}$ is the vector of Pauli matrices, acting in the HH subspace.

\subsubsection{Spherical approximation, $\delta=0$}

Let us again first review the case where the spherical approximation $\delta \to 0$ is reasonable, such as for Ge.
At the edge of the valence band, i.e., where $p_x=p_y=0$, we have $R=S=0$ and the two HH states are thus pure $m_j = \pm \frac{3}{2}$ states.
In that case, the effective HH Zeeman Hamiltonian becomes to leading order in $1/u_z$
\begin{align}
H_{\rm Z}^{\rm H} = 3\kappa B_z\sigma_z.
\end{align}
At the edge of the valence band, the coupling to the in-plane components of the magnetic field $B_{x,y}$ is a higher-order effect via the LH states and is thus proportional to $B^3_{x,y}/u_z^2$.
In terms of the $g$-tensor this means that $g_{zz} = 6\kappa$ and all other elements are much smaller.

For a 2DHG with a finite density, the holes with non-zero in-plane momentum have a non-zero matrix element $R$, see Eq.~(\ref{eq:R_spherical}).
This means that for holes close to the Fermi level the resulting HH-LH mixing adds a finite coupling to the in-plane field, yielding an effective direction-dependent $g$-tensor
\begin{align}
\bar g = \left( \begin{array}{ccc}
g_\parallel \cos 2\varphi & -g_\parallel \sin 2\varphi & 0 \\
g_\parallel \sin 2\varphi & g_\parallel \cos 2\varphi & 0 \\
0 & 0 & g_\perp
\end{array}\right),
\end{align}
with $g_\perp = 6\kappa$ and $g_\parallel = 6\kappa p_{\rm F}^2/2m_0u_z$, again up to order $\mathcal{O}(1/u_z)$. Here $p_{\rm F}$ is the Fermi momentum and $\varphi$ is the direction of propagation of the hole under consideration.

Using that we defined $u_z = \langle p_z^2 \rangle / 2m_0$, we arrive at an elegant expression for the ratio of the magnitudes of the in-plane and out-of-plane $g$-factors in the spherical approximation~\cite{winkler2003,Marcellina2018},
\begin{equation}
\frac{g_\parallel}{g_\perp} = \frac{p_{\rm F}^2}{\langle p_z^2 \rangle}.
\end{equation}
Assuming parabolic dispersion for the range of energies of interest, we can consider a finite two-dimensional density of HHs $\rho$ in the valence band and thus write for the ratio of $g$-factors at the Fermi level
\begin{equation}
\frac{g_\parallel}{g_\perp} = \frac{2\pi \rho}{\langle k_z^2 \rangle} = \frac{2}{\pi}\rho d^2,
\end{equation}
where in the last step we again used our assumption of an infinite-well type of confinement along $z$, resulting in $\langle k_z^2 \rangle = \pi^2/d^2$, where $d$ is the width of the well.

\subsubsection{Anisotropic Hamiltonian, $\delta\neq 0$}

Going beyond the spherical approximation, as is necessary for Si, all HHs and LHs are mixtures of $m_j = \pm\frac{3}{2}$ and $m_j = \pm \frac{1}{2}$ states, thus resulting in general in a finite coupling to $B_{x,y}$ within the HH subspace, also in the absence of finite in-plane momentum.
We thus transform the Zeeman Hamiltonian (\ref{eq:zeeman}) to the basis where the part of $H$ proportional to $u_z$ is diagonal, which we then project to the HH subspace.
To leading order in $1/u_z$, the resulting $g$-tensor can be written relatively compactly,
\begin{align}
    \frac{g_{zz}}{\kappa} &= 2\frac{Q_{zz}}{\nu} + 4\frac{\nu}{\mu},
    \label{eq:g-tensor_start}
    \\
    \frac{g_{zx}-ig_{zy}}{\kappa} &= 2\sqrt{3}\frac{S_{zz}}{\nu} - 2\frac{R_{zz}S^*_{zz}}{\mu\nu} ,
    \\
    \frac{g_{xz}+ig_{yz}}{\kappa} &= 2\frac{R_{zz}S_{zz}}{\mu\nu},
    \\
    g_{xx}-ig_{xy} &= g_{\scaleto{-+}{4pt}} + g_{\scaleto{++}{4pt}},
    \\
    g_{yy}+ig_{yx} &= g_{\scaleto{-+}{4pt}} - g_{\scaleto{++}{4pt}},
    \label{eq:g-tensor_end}
\end{align}
with 
\begin{align*}
    \frac{g_{\scaleto{-+}{4pt}}}{\kappa} &=
    -\sqrt{3}\frac{R_{zz}^*}{\mu}\left(1+\frac{Q_{zz}}{\nu}\right)
    \\&\phantom{=}\ 
    -\frac{(S_{zz}^*)^2}{|S_{zz}|^2}\left(1-\frac{Q_{zz}}{\nu}\right)\left(1+\frac{\nu}{\mu}\right),
    \\
    \frac{g_{\scaleto{++}{4pt}}}{\kappa} &= 
    \sqrt{3}\frac{R_{zz}}{\mu}\frac{S_{zz}^2}{|S_{zz}|^2}\left(1-\frac{Q_{zz}}{\nu}\right)
    \\&\phantom{=}\ 
    + \frac{R_{zz}^2}{|R_{zz}|^2}\left(1+\frac{Q_{zz}}{\nu}\right)\left(1-\frac{\nu}{\mu}\right),
    \hspace{-0.1cm} 
\end{align*}
using the shorthand notation $\nu=\sqrt{Q_{zz}^2+|S_{zz}|^2}$ and $\mu=\sqrt{Q_{zz}^2+|S_{zz}|^2+|R_{zz}|^2}$. 
This result is again valid to leading order in $1/u_z$; we note that with the spherical approximation we have $S_{zz}=R_{zz}=0$, yielding $g_{zz} = 6\kappa$ as only non-zero element, as expected~\footnote{Although we here focus on the leading order terms $\propto 1/u_z$, we note that the expressions in Eqs.~\eqref{eq:g-tensor_start}-\eqref{eq:g-tensor_end} can be generalized to describe the $g$-tensor of any heavy hole governed by a Hamiltonian in the form of Eq.~\eqref{eq:Ham_mmn} by simply substituting $\{Q_{zz},R_{zz},S_{zz}\}\rightarrow\{Q,R,S\}$.}.
These analytic results generalize those presented in Refs.~\cite{Winkler2000,Winkler2008}, where the focus was on confinement along $[nnm]$.

When the 2DHG is confined along a high-symmetry direction, like [001] or [111], we obtain an out-of-plane $g$-factor $g_\bot = 6\kappa$ and in-plane $g$-factors $g_\parallel=0$, as expected.
For lower-symmetry directions also the $g_{xx}$ and $g_{yy}$ components become non-zero, and by taking [110] as an example we find straightforwardly
\begin{align}
    \frac{g_{zz}}{\kappa} &= 2+\frac{2(\gamma_2+3\gamma_3)}{\sqrt{\gamma_2^2+3\gamma_3^2}},
    \\
    \frac{g_{yy}}{\kappa} &= -2 - \frac{2(\gamma_2 - 3\gamma_3)}{\sqrt{\gamma_2^2+3\gamma_3^2}},
    \\
    \frac{g_{xx}}{\kappa} &= 2-\frac{ 4\gamma_2}{\sqrt{\gamma_2^2+3\gamma_3^2}},
\end{align}
and vanishing off-diagonal components.
To obtain non-zero off-diagonal elements one has to consider less common directions.
For example, for a 2DHG oriented along [112] one finds non-zero off-diagonal elements
\begin{align}
     \frac{g_{xz}}{\kappa} &= \frac{2\sqrt{2}}{\sqrt{3}}\frac{(\gamma_3-\gamma_2)^2}{\sqrt{\gamma_2^2+3\gamma_3^2} \sqrt{11\gamma_2^2+2\gamma_2\gamma_3+35\gamma_3^2}},
     \\
     \frac{g_{zx}}{\kappa} &= \frac{2\sqrt{2}}{\sqrt{3}}
     \frac{6(\gamma_3-\gamma_2)\sqrt{\gamma_2^2+3\gamma_3^2}-(\gamma_3-\gamma_2)^2}{\sqrt{\gamma_2^2+3\gamma_3^2} \sqrt{11\gamma_2^2+2\gamma_2\gamma_3+35\gamma_3^2}},
\end{align}
coupling the $z$-component of the spin to the $x$-component of $\bs B$ and vice-versa, making the $g$-tensor highly anisotropic.
All these expressions follow straightforwardly from Eqs.~(\ref{eq:g-tensor_start}-\ref{eq:g-tensor_end}) upon inserting the explicit expressions given in App.~\ref{app:Luttinger_elements}.

\begin{figure}[t]
	\centering
	\includegraphics[width=.95\linewidth]{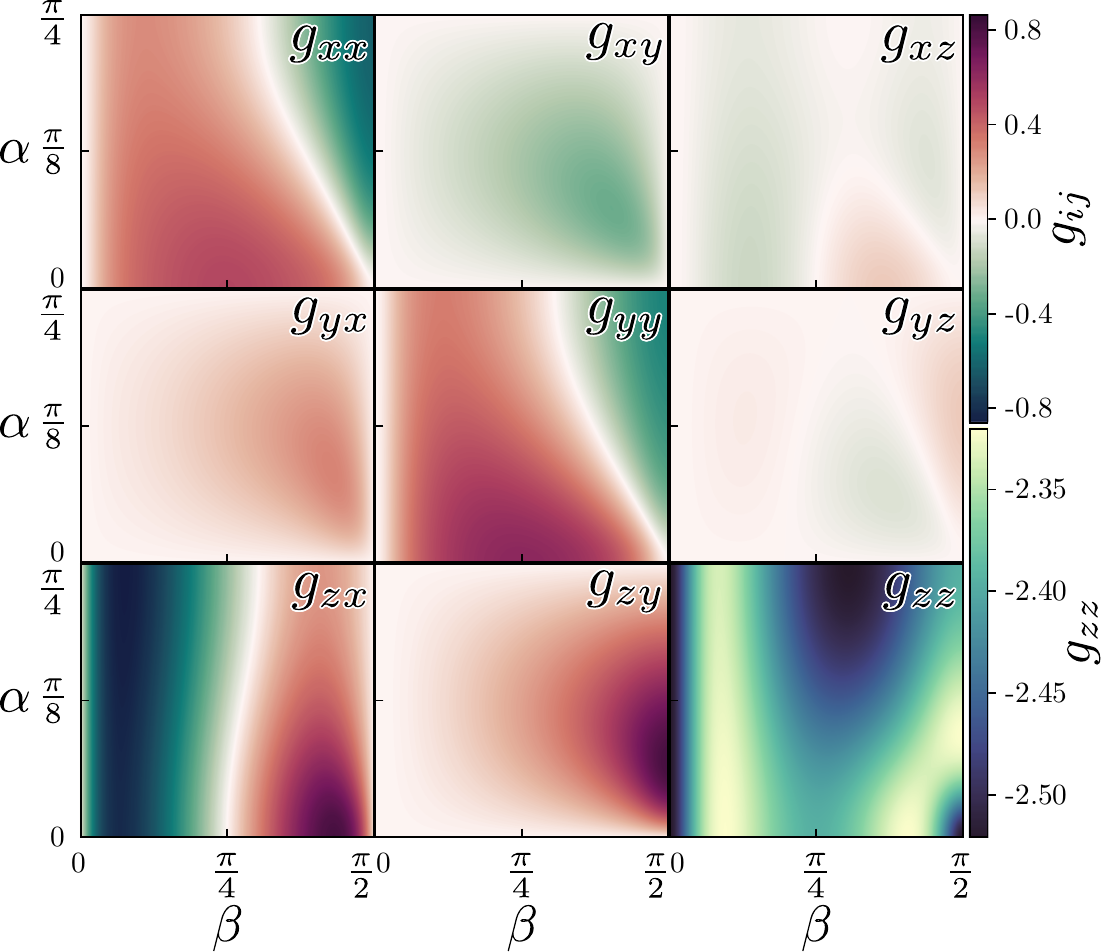} 
	\caption{The nine components of the heavy hole $g$-tensor given by Eqs.~\eqref{eq:g-tensor_start}-\eqref{eq:g-tensor_end} plotted against the direction of confinement. We here used parameters for Si, see Tab.~\ref{tab:Luttinger_parameters}. }
	\label{fig:g-tensor_elements}
\end{figure}
In Fig.~\ref{fig:g-tensor_elements} we plot the magnitude of all nine components of the HH $g$-tensor at the band edge as a function of the two confinement angles $\alpha$ and $\beta$, as given by Eqs.~(\ref{eq:g-tensor_start})--(\ref{eq:g-tensor_end}), where we again used parameters for Si. 
We see that by controlling the orientation of the confining potential one can design the qualitative form of the $g$-tensor, ranging from purely diagonal for high-symmetry directions to highly anisotropic for less common directions.

\section{Confined heavy holes}
\label{sec:Quantum_dots}

In the previous section we investigated the effects of confinement along the $z$-direction on the $g$-tensor in a 2DHG.
Further confinement along the in-plane coordinates $x$ and $y$, often done via electrostatic gating, can then be used to localize the holes in lateral quantum dots, opening up the possibility to use them as a spin-qubit platform.
The effective $g$-tensor for such localized holes can be affected by spin--orbit interaction (SOI)~\cite{Hanson2007}, the effect of which we will include in this section.

In this section we will restrict ourselves to a general linear Rashba-type SOI, which could be caused by the 2DHG confinement potential~\cite{Nakamura2012,Moriya2014}.
The Hamiltonian describing this type of interaction for the $j=\frac{3}{2}$ states in the upper valence band reads as~\cite{WinklerSOI2000,winkler2003,Winkler2008}
\begin{equation}
    H_\text{so} = \beta_\mathrm{so} (p_y J_x - p_x J_y),
    \label{eq:SOI_isotropic}
\end{equation}
where we neglected the contribution proportional to $\mathcal{J}$, which is usually much weaker, and we assumed the electric field associated with the confining potential to point along $z$.
The parameter $\beta_\mathrm{so}$ is material-dependent and depends also in an intricate way on the exact shape of the transverse confining potential.
By focusing solely on this Rashba term, we neglect the Dresselhaus contribution stemming from the lack of a crystallographic inversion center (which can contribute to the SOI in materials like GaAs and InAs) and we disregard the ``direct'' Rashba SOI due to HH-LH mixing~\cite{Kloeffel2011,Kloeffel2018,Bosco2021a,Bosco2021,Xiong2021}.
With this choice, the analysis that follows is most relevant for quantum dots hosted in materials with a large Rashba coefficient.
However, our approach can straightforwardly be adapted to other types of SOI, as we will indicate below.

In the remainder of this section we will start by calculating the level structure of holes confined in a quantum dot.
This allows us then to project the spin--orbit Hamiltonian in Eq.~\eqref{eq:SOI_isotropic} to this basis of localized heavy-hole states and calculate the SOI-induced corrections to the $g$-tensor using perturbation theory.

\subsection{Level structure}

The Luttinger Hamiltonian that governs the in-plane motion of the effective heavy holes was obtained by transforming the in-plane part of $H$ in Eq.~\eqref{eq:Ham_mmn} to the basis where the part of $H$ proportional to $u_z$ is diagonal.
We now add a circularly symmetric parabolic confinement potential $V({\bf r}) = \lambda (x^2+y^2)$, which can describe the confinement of the holes in a quantum dot,
\begin{equation}\label{eq:hamdot}
    H_{\text{L},\parallel}^\text{H} = \frac{p_{\tilde x}^2}{2m_-} + \frac{p_{\tilde y}^2}{2m_+} + \frac{m_-}{2}\omega_x^2 \tilde x^2+\frac{m_+}{2}\omega_y^2 \tilde y^2.
\end{equation}
Here, $m_-=m^{{\rm H}}(\zeta)$ and $m_+=m^{{\rm H}}(\zeta+\pi/2)$ are the minimum and maximum HH effective masses, as given by Eq.~(\ref{eq:masses}), and the new in-plane coordinate system $\{\tilde x, \tilde y \}$ is thus rotated over an angle $\zeta$ along $z$ with respect to the original system $\{x,y\}$.
Further, the frequencies $\omega_x = \sqrt{2\lambda/m_-}$ and $\omega_y = \sqrt{2\lambda/m_+}$ determine the strength of the in-plane confinement and ${\bs p} = -i\hbar\partial_{\tilde{\bs r}}+e\bs{A}(\tilde{\bs{r}})$ is the canonical momentum, with $\bs{A}(\tilde{\bs{r}}) = B_z(-\tilde y/2,\tilde x/2,0)$ being the vector potential for which we use the circular gauge and neglect in-plane components of the magnetic field, assuming strong confinement along $z$.

The eigenstates and -energies of such an anisotropic two-dimensional oscillator in the presence of a magnetic field can be found in different ways, see e.g.~Refs.~\cite{Rebane1972,Schuh1985,Froning2021}. We follow the procedure presented in \cite{Qiong_Gui2002}, resulting in a Hamiltonian that can be written in terms of two independent harmonic oscillators,
\begin{equation}\label{eq:ham_in-plane}
    H_{\text{L},\parallel}^\text{H} =
    \hbar\omega_+\left(a_+^\dagger a_+ +\frac{1}{2}\right) +\hbar\omega_-\left(a_-^\dagger a_- +\frac{1}{2}\right),
\end{equation}
where $a_\pm^{(\dagger)}$ are bosonic creation and annihilation operators, and the (positive) oscillator frequencies are defined through
\begin{align}
    \omega_\pm^2 = {} & {} \frac{1}{2}\omega_x^2+\frac{1}{2}\omega_y^2+2\omega_c^2 
    \nonumber\\
    {} & {} \pm \frac{1}{2}\sqrt{(\omega_x^2-\omega_y^2)^2 + 8\left( \omega_x^2+\omega_y^2 + 2\omega_c^2 \right)\omega_c^2},
\end{align}
with $\omega_c^2 = e^2B_z^2/4m_+m_-$.
We will assume throughout that the oscillator energies $\hbar\omega_\pm$ are much smaller than the energy $u_z$ associated with the transverse confinement.

Since the masses $m_\pm$ depend on the orientation of the plane of the 2DHG (through the angles $\alpha$ and $\beta$, see Sec.~\ref{sec:2DHG_beyond}) the level splitting in the dot will also vary as a function of that orientation, which can be shown more explicitly by inserting the maximum and minimum masses as given by Eq.~(\ref{eq:masses}),
\begin{align}
    \omega^2_\pm = \frac{\lambda}{m_0} \big( 2\gamma_1 + a+2\chi_c^2 
    \pm \eta\big),
    \label{eq:level_splitting}
\end{align}
where we introduced the notation
\begin{equation}
    \eta = \sqrt{b^2 + 4\chi_c^2(2\gamma_1+a+\chi_c^2) },
\end{equation}
and used the dimensionless parameter
\begin{equation}
\chi_c = \frac{eB_z}{4\sqrt{\lambda m_0}}\sqrt{ (2\gamma_1 + a)^2 - b^2},
\end{equation}
characterizing the magnitude of the cyclotron frequency compared to the harmonic oscillator frequencies.
We used the same notation as in Sec.~\ref{sec:2DHG_beyond}, where we omitted the superscript H from the coefficients $a$ and $b$.

\begin{figure}[tb]
\centering
\includegraphics[width=\linewidth]{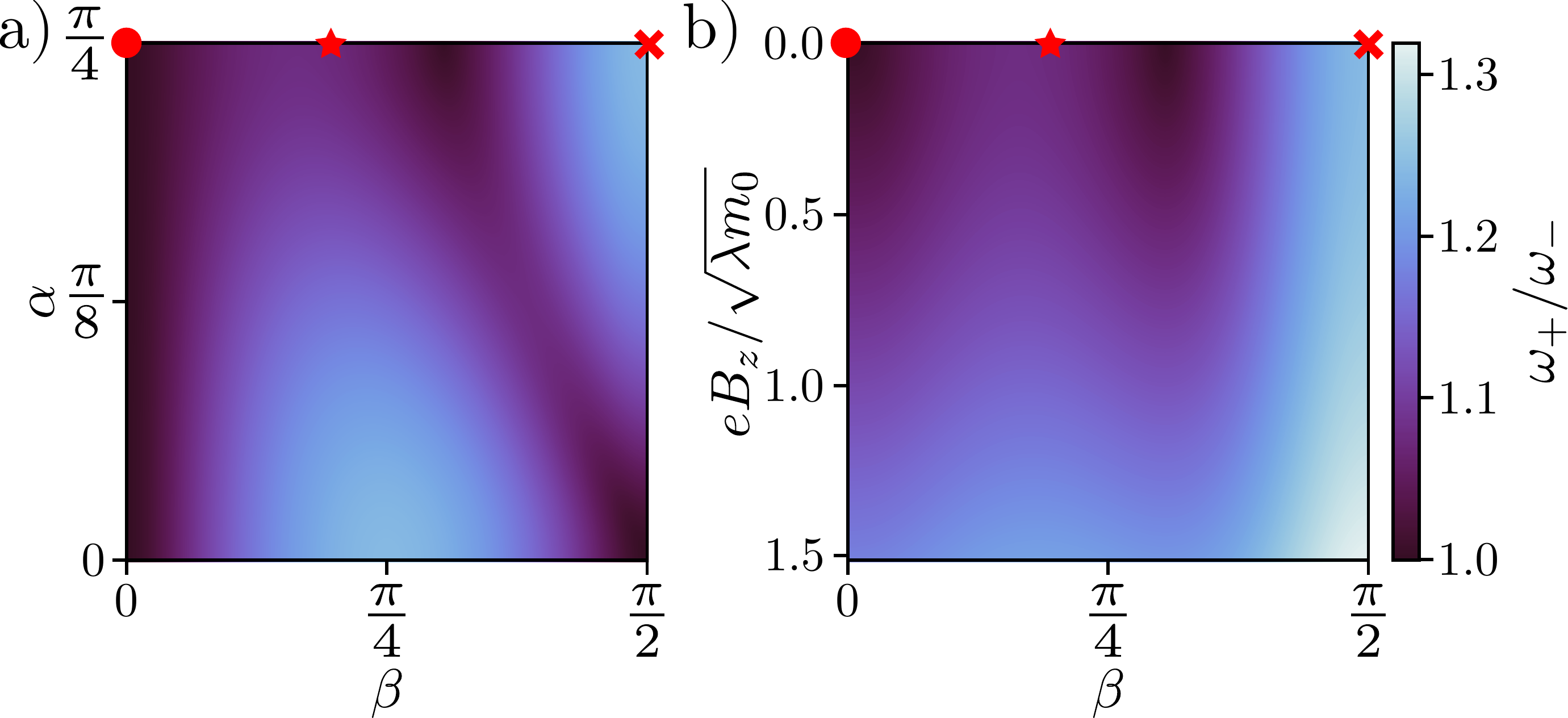} 
\caption{Asymmetry in the confinement energy $\omega_+/\omega_-$ as given by Eq.~\eqref{eq:level_splitting}. (a) $\omega_+/\omega_-$ as a function of the direction of confinement in the absence of a vector potential. (b) $\omega_+/\omega_-$ as a function of the angle $\beta$ and applied magnetic field $eB_z/\sqrt{\lambda m_0}$, setting $\alpha=\pi/4$ (which corresponds to focusing on confinement directions $[nnm]$). In both plots we used parameters for Si.}
\label{fig:level_splitting}
\end{figure}

To illustrate the dependence of the confinement energies on the orientation of the 2DHG explicitly, we plot in Fig.~\ref{fig:level_splitting}(a) the anisotropy of the level splitting $\omega_+/\omega_-$ as a function of the angles $\alpha$ and $\beta$ in the absence of a vector potential, where we used parameters for Si.
Naturally, since this anisotropy stems from the orientation dependence of the effective mass, it strongly resembles the results shown in Fig.~\ref{fig:effective_masses}(c).
Fig.~\ref{fig:level_splitting}(b) shows how a non-zero vector potential affects the anisotropy.
We plot $\omega_+/\omega_-$ as a function of $eB_z/\sqrt{\lambda m_0}$ and the angle $\beta$, fixing $\alpha = \pi/4$, which captures all confinement directions of the form $[nnm]$.
We see that, as expected, the magnetic field increases the anisotropy, while retaining some of the orientation dependence.

\subsection{Corrections to the $g$-tensor}

Since the holes we now consider are localized, $H_\text{so}$ does not couple the orbital ground states of the heavy holes directly, but does so only in higher order via virtually excited orbital states, this in contrast with the Zeeman Hamiltonian.
To find the corrections to the $g$-tensor for the localized eigenstates of the Hamiltonian (\ref{eq:hamdot}), we first transform $H_{\rm so}$ to the basis that diagonalizes the part of the Luttinger Hamiltonian (\ref{eq:Ham_mmn}) that is proportional to $u_z$, as we did in Sec.~\ref{sec:g-tensor}.
This transformation thus amounts to performing the same rotation as we applied to $H_{\rm Z}$, meaning that the resulting Hamiltonian in the HH subspace can be written using the $g$-tensor we derived above.
For a general spin--orbit Hamiltonian ${\bs f}(p_x,p_y) \cdot {\bs J}$ this results in $\frac{1}{4\kappa}{\bs \sigma}\cdot \bar g\cdot {\bs f}(p_x,p_y)$, and for the case of the linear Rashba Hamiltonian (\ref{eq:SOI_isotropic}) we thus write
\begin{equation}
    H_\mathrm{so}^\mathrm{H} = \frac{\beta_\mathrm{so}}{4\kappa} \bs\sigma\cdot\bar g\cdot \left( \begin{array}{c} p_y \\ -p_x \\ 0 \end{array} \right).
    \label{eq:soi_hh}
\end{equation}
In this form the spin--orbit Hamiltonian contains the leading-order effect of HH-LH mixing.

We then express the in-plane momentum operators $p_{x,y}$ as linear combinations of the bosonic creation and annihilation operators $a_\pm^\dagger$ and $a_\pm$ (see Appendix \ref{app:Level_structure}),
\begin{align}
    p_x &= w_x^+a_+ + w_x^-a_- + \text{H.c.},
    \label{eq:momentum_sub1}
    \\
    p_y &= w_y^+a_+ + w_y^-a_- + \text{H.c.},
    \label{eq:momentum_sub2}
\end{align}
with
\begin{align}
    w_x^\pm = {} & {} W^\mp \bigg[\pm\left(\frac{m_-^3}{m_+}\right)^{\!\! 1/8}\!\! \sqrt{\eta\pm b \pm 2\chi_c^2} \cos\zeta 
    \nonumber\\ {} & {} \hspace{3em}
    -i \left(\frac{m_+^3}{m_-}\right)^{\!\! 1/8}\!\! \sqrt{\eta\mp b \pm 2\chi_c^2}  \sin\zeta \bigg],\\
    w_y^\pm = {} & {} W^\mp \bigg[i\left(\frac{m_+^3}{m_-}\right)^{\!\! 1/8}\!\! \sqrt{\eta\mp b \pm 2\chi_c^2} \cos\zeta 
    \nonumber\\ {} & {} \hspace{3em}
    \pm \left(\frac{m_-^3}{m_+}\right)^{\!\! 1/8}\!\! \sqrt{\eta\pm b \pm 2\chi_c^2}  \sin\zeta \bigg],
\end{align}
where the common prefactor reads as
\begin{equation}
    W^\pm = \left(\frac{\hbar^2 \lambda}{16\eta^2}\right)^{\!\! 1/4} \! \left[(2\gamma_1 + a +2 \chi_c^2-\eta) \sqrt{\frac{m_+m_-}{m_0^2}}\right]^{\! \pm 1/4},
\end{equation}
and the angle $\zeta$ was defined in (\ref{eq:angle_zeta}).

By inserting Eqs.~\eqref{eq:momentum_sub1} and \eqref{eq:momentum_sub2} for $p_x$ and $p_y$ in the spin--orbit Hamiltonian (\ref{eq:soi_hh}) we see that we can write
\begin{equation}
    H_\mathrm{so}^\mathrm{H} =  \frac{\beta_{\rm so}}{4\kappa} \sigma_\alpha A_\alpha^\gamma a_\gamma + \text{H.c.},
    \label{eq:SOI_ladders}
\end{equation}
in terms of bosonic creation and annihilation operators, which makes performing perturbation theory in the SOI very straightforward.
Summation over repeated indices $\alpha \in \{x,y,z\}$ and $\gamma \in \{ +,-\}$ is implied and we introduced the vectors
\begin{equation}
    A_{\alpha}^{\pm} =  g_{\alpha x}w^\pm_y - g_{\alpha y}w^\pm_x.
\end{equation}
Similar expressions for other types of SOI can straightforwardly be derived along the same lines.

\begin{figure}[t]
	\centering
	\includegraphics[width=.95\linewidth]{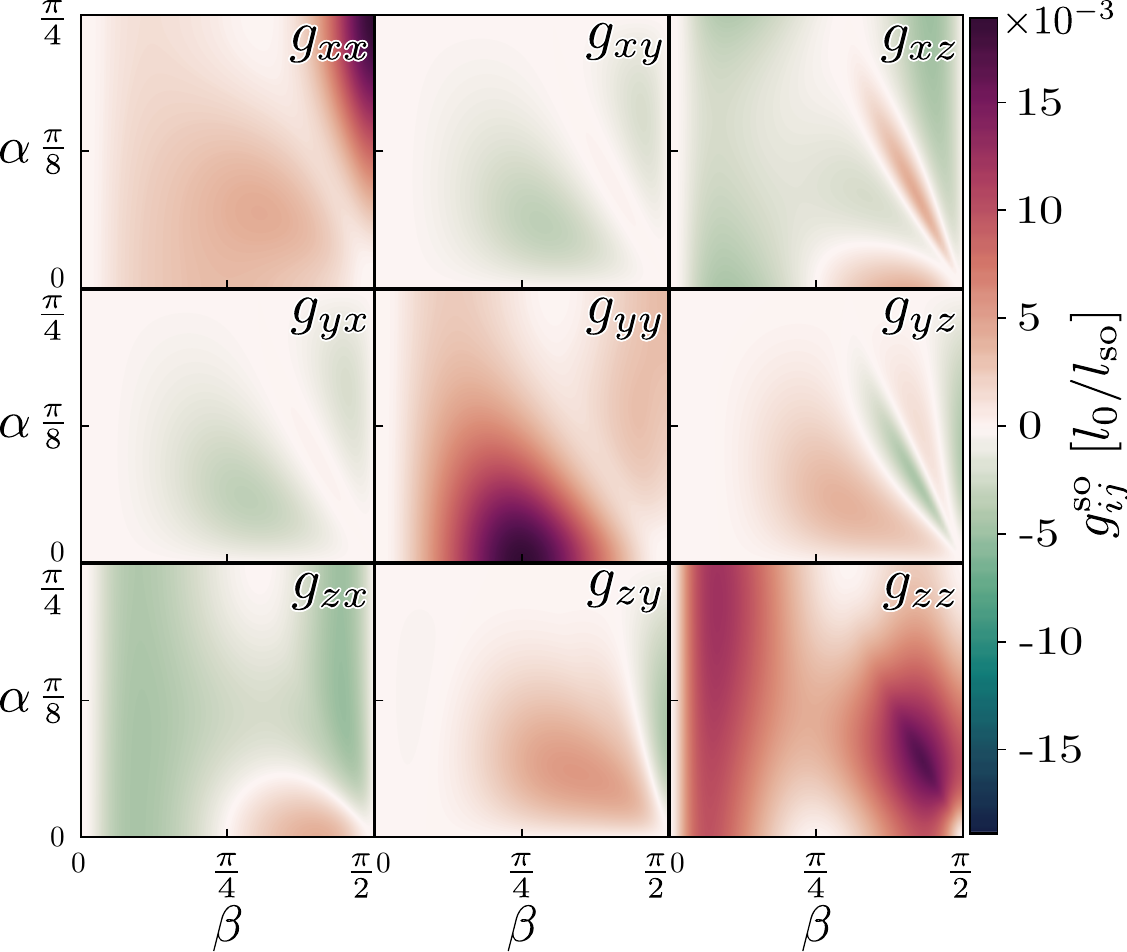} 
	\caption{The nine matrix elements of the spin--orbit correction to the $g$-tensor, as given by Eq.~\eqref{eq:g_soi}, plotted as a function of the direction of the confinement plane. The correction is shown in units of $l_0/l_\mathrm{so}$. In this plot we used again parameters for Si.}
	\label{fig:g-tensor_elements_so}
\end{figure}

Using this form of the spin--orbit Hamiltonian we now perform second-order perturbation theory on the eigenstates of the unperturbed Hamiltonian $H_0 = H^{\rm H}_{\text{L},\parallel} + H^{\rm H}_{\rm Z}$, as given in Eqs.~(\ref{eq:Zeeman_HH}) and (\ref{eq:ham_in-plane}).
To order $\beta_{\rm so}^2/\omega_\pm^2$ this yields a correction that follows from projecting
\begin{equation}
V_{\rm so} = \sum_{\gamma=\pm} \big\{
H_{\rm so}^{\rm H} [ \gamma \tfrac{1}{2} E_{\rm Z} - H_0 ]^{-1} H_{\rm so}^{\rm H}, P_\gamma \big\},
\label{eq:SOI_g}
\end{equation}
to the HH subspace.
Here $E_Z = |\bar{g} \cdot \bs{B}|$ is the magnitude of the HH Zeeman splitting, the operator $P_\pm = |\pm\rangle\langle\pm|$ projects to the two eigenstates of $H^{\rm H}_{\rm Z}$ and can be written explicitly as $P_\pm = \frac{1}{2}(1 \pm \bs b \cdot \bs \sigma)$ with $\bs b = \bar g \cdot \bs B / E_{\rm Z}$ the unit vector pointing in the direction of the Zeeman field. 
We then evaluate \eqref{eq:SOI_g} and extract the spin--orbit-induced contribution to the $g$-tensor from the linear dependence of $V_{\rm so}$ on ${\bs B}$.
This yields the relatively compact expression
\begin{align}
    g^\mathrm{so}_{ij} =  \frac{1}{16\kappa^2}\frac{l_0^2}{l_\mathrm{so}^2}
    \bigg[ {} & {} 
    \cos^2\zeta\left( \frac{g_{ix}g_{jx}}{L_-^3} + \frac{g_{iy}g_{jy}}{L_+^3} \right) 
    \nonumber\\{} & {} 
    + \sin^2\zeta\left( \frac{g_{ix}g_{jx}}{L_+^3} + \frac{g_{iy}g_{jy}}{L_-^3} \right)
    \nonumber\\{} & {} 
    + \sin(2\zeta)\frac{L_++L_-}{L_+^2L_-^2} \epsilon_{ikl}g_{kx}g_{ly} \delta_{jz} \bigg],
    \label{eq:g_soi}
\end{align}
where
\begin{equation}
    L_\pm = \sqrt{2\gamma_1+a\pm b},
\end{equation}
and we used the length scales $l_0 = (\hbar^2/m_0\lambda)^\frac{1}{4}$ (characterizing the in-plane confinement) and $l_\mathrm{so} = \hbar/m_0\beta_\mathrm{so}$ (the spin--orbit length).
The first two terms in (\ref{eq:g_soi}) arise due to the Zeeman shift of the ground and excited spin states, whereas the last term contains the contribution linear in $\omega_c$ and couples therefore only to $B_z$.

In Fig.~\ref{fig:g-tensor_elements_so} we show an example of the orientation dependence of the matrix elements $g_{ij}^\mathrm{so}$ as given by Eq.~\eqref{eq:g_soi}, where we again used parameters for Si, for consistency.
The matrix elements are plotted in units of the dimensionless ratio $l_0/l_\mathrm{so}$, which characterizes the effect of the spin--orbit interaction in the quantum dots.
The elements $g^\mathrm{so}_{ix}$ and $g^\mathrm{so}_{iy}$ are solely determined by the first two terms in Eq.~\eqref{eq:g_soi}, whereas the elements $g^\mathrm{so}_{iz}$ also include contributions from the last term.
Similar to the unperturbed $g$-tensor as investigated in Sec.~\ref{sec:g-tensor}, many elements of the spin--orbit correction $\bar g^{\mathrm{so}}$ also vanish for high-symmetry confinement directions such as [001] and [111].

Depending on the details of the material and confinement potential of the hole gas, other types of SOI than the linear Rashba type of Eq.~(\ref{eq:soi_hh}) could be dominating, such as an effectively cubic Rashba interaction $\propto p_+^3\sigma_- - p^3_-\sigma_+$.
We emphasize again that the derivation presented in this section can easily be adapted to such other spin--orbit Hamiltonians, simply by substituting Eqs.~(\ref{eq:momentum_sub1},\ref{eq:momentum_sub2}) into the spin--orbit Hamiltonian and evaluating the resulting correction (\ref{eq:SOI_g}).
Working in the bosonic number basis this is a straightforward task.

\section{Conclusion}
Depending on the choice of material, holes confined in two- or lower-dimensional semiconductor structures can possess anisotropic dynamics that are highly dependent on the details of the  confinement potentials in the system.
Such holes can have several interesting properties that arise from this anisotropy, such as highly anisotropic effective masses and \textit{g}-tensors.

In this paper we investigated these anisotropies, with special focus on the detailed role of the orientation of the confinement potentials.
Starting from a $4\times 4$ Luttinger Hamiltonian, which we did not necessarily assume to be spherically symmetric, we assumed very strong transverse confinement in one direction, resulting in a 2DHG.
We rotated our coordinate system such that the transverse direction could easily be integrated out for an arbitrary direction of confinement.
This approach allowed us to extract very general analytic expressions for both the in-plane effective hole masses and the heavy-hole \textit{g}-tensor, where we pointed out how the effect of strain can easily be included.
We then investigated a strainless 2DHG and derived analytic expressions for the effective masses and the $g$-tensor.
In our explicit results we focused on Si, which exhibits relatively strong anisotropies, but the expressions we presented are fully general.

We then assumed additional in-plane confinement, leading to the formation of quantum dots.
We presented a straightforward approach to include the effects of spin--orbit coupling on the dynamics of the hole states localized in a quantum dot.
As an example we considered the effect of a linear Rashba-like SOI that could arise from the transverse confinement.
By calculating the level splitting of the localized states, we projected the spin--orbit Hamiltonian to the basis of the localized states and used perturbation theory to obtain an electric-field-dependent correction to the \textit{g}-tensor of the confined heavy holes.
Our results are highly relevant for the ongoing efforts to use hole spins localized in Si- or Ge-based quantum dots as spin qubits.
Finding optimal working points, providing fast qubit control through $g$-tensor modulation together with relative insensitivity to charge noise requires a thorough understanding of the intricate interplay of SOI, confinement, and applied magnetic fields.

\appendix
\section{Hamiltonian tensor elements}
\label{app:Luttinger_elements}

The rotated Luttinger (and Bir-Pikus) Hamiltonian can always be written in the following form,
\begin{equation}
H(\alpha,\beta) = \left(\begin{array}{cccc}
P-Q & -S & R & 0 \\
-S^\dagger & P+Q & 0 & R \\
R^\dagger & 0 & P+Q & S \\
0 & R^\dagger & S^\dagger & P-Q
\end{array}\right),
\end{equation}
in the basis of the eigenstates $\{\ket{\frac{3}{2}},\ket{\frac{1}{2}},\ket{-\frac{1}{2}},\ket{-\frac{3}{2}}\}$ of $J_z$ with its quantization axis along the new $z$-direction.

For the Luttinger Hamiltonian the matrix elements $P$, $Q$, $R$ and $S$ can be expressed in terms of dimensionless symmetric tensors $M_{ij}$,
\begin{equation}
M = \frac{1}{2m_0}\sum_{i,j}M_{ij}\{ p_i, p_j \},
\end{equation}
where $\{A,B\}=\frac{1}{2}\left(AB+BA\right)$, $M\in\{P,Q,R,S\}$ and $i,j\in\{x,y,z\}$.
The diagonal element $P$ is invariant under rotations and follows from $P_{ij} = \delta_{ij}\gamma_1$, while the tensor elements of $Q$, $R$ and $S$ read
\begin{widetext}
\begin{align}
    Q_{xx} &= -\frac{1}{5}\left(2\gamma_2+3\gamma_3\right) - \frac{3\delta}{160}\left[3+5\cos(4\alpha)-5\cos(4\beta)\left\{7+\cos(4\alpha)\right\}\right],
    \\
    Q_{yy} &= -\frac{1}{5}\left(2\gamma_2+3\gamma_3\right) + \frac{3\delta}{40}\left[3+5\cos(4\alpha)+10\cos(2\beta)\sin^2(2\alpha)\right],
    \\
    Q_{zz} &= \frac{2}{5}\left(2\gamma_2+3\gamma_3\right) - \frac{3\delta}{160}\left[9+20\cos(2\beta)+35\cos(4\beta)+40\cos(4\alpha)\sin^4\beta\right],
    \\
    Q_{xy} &= -\frac{3\delta}{2}\sin(4\alpha)\cos\beta\sin^2\beta,
    \\
    Q_{yz} &= -\frac{3\delta}{2}\sin(4\alpha)\sin^3\beta,
    \\
    Q_{zx} &= \frac{3\delta}{16}\left[4\sin^2(2\alpha)\sin(2\beta)+\sin(4\beta)\left\{7+\cos(4\alpha)\right\}\right],\\
    R_{xx} &= -\frac{\sqrt{3}}{5}\left(2\gamma_2+3\gamma_3\right) + \frac{\sqrt{3}\delta}{160}\left[ 21-40\cos(2\beta)+35\cos(4\beta)+80\cos^2\beta\left\{e^{-4i\alpha}\cos^4\left(\tfrac{\beta}{2}\right)+e^{4i\alpha}\sin^4\left(\tfrac{\beta}{2}\right)\right\} \right],
    \\
    R_{yy} &= \frac{\sqrt{3}}{5}\left(2\gamma_2+3\gamma_3\right)-\frac{\sqrt{3}\delta}{40}\left[9+15\cos(4\alpha)-10\sin^2(2\alpha)\cos(2\beta)-20i\sin(4\alpha)\cos\beta\right],
    \\
    R_{zz} &= \frac{\sqrt{3}\delta}{8}\sin^2\beta\left[ 5+7\cos(2\beta)+\cos(4\alpha)\left\{3+\cos(2\beta)\right\}-4i\sin(4\alpha)\cos\beta \right],
    \\
    R_{xy} &= \frac{2\sqrt{3}i}{5}\left(2\gamma_2+3\gamma_3\right) + \frac{\sqrt{3}\delta}{40}\left[5\sin(4\alpha)\left\{7\cos\beta+\cos(3\beta)\right\} + 4i\left\{3+5\cos(4\alpha)-10\sin^2(2\alpha)\cos(2\beta)\right\}\right],
    \\
    R_{yz} &= \frac{\sqrt{3}\delta}{8}\left[\sin(4\alpha)\left\{5\sin\beta+\sin(3\beta)\right\}-8i\sin^2(2\alpha)\cos(2\beta)\right],
    \\
    R_{zx} &= -\frac{\sqrt{3}\delta}{8}\sin(2\beta)\left[3-7\cos(2\beta)-\cos(4\alpha)\left\{3+\cos(2\beta)\right\}+4i\sin(4\alpha)\cos\beta\right],\\
    S_{xx} &= -\frac{\sqrt{3}\delta}{16}\left[8\cos^2\beta\sin\beta\left\{\cos(4\alpha)\cos\beta-i\sin(4\alpha)\right\} - 2\sin(2\beta) + 7\sin(4\beta)\right],
    \\
    S_{yy} &= -\sqrt{3}\delta\sin(2\alpha)\sin\beta\left[\sin(2\alpha)\cos\beta+i\cos(2\alpha)\right],
    \\
    S_{zz} &= -\frac{\sqrt{3}\delta}{16}\left[8\sin^3\beta\left\{\cos(4\alpha)\cos\beta-i\sin(4\alpha)\right\}-2\sin(2\beta)-7\sin(4\beta)\right],
    \\
    S_{xy} &= -\sqrt{3}\delta\sin(2\alpha)\sin(2\beta) \left[\cos(2\alpha)\cos\beta-i\sin(2\alpha)\right],
    \\
    S_{yz} &= -\frac{2\sqrt{3}i}{5}\left(2\gamma_2+3\gamma_3\right) - \frac{\sqrt{3}i\delta}{10}\left[3+5\cos(2\beta)+10\sin^2\beta\left\{\cos(4\alpha)-i\sin(4\alpha)\cos\beta\right\}\right],
    \\
    S_{zx} &= \frac{2\sqrt{3}}{5}\left(2\gamma_2+3\gamma_3\right) - \frac{\sqrt{3}\delta}{40}\left[3+5\cos(4\alpha)\left\{1-\cos(4\beta)\right\}-35\cos(4\beta) - 40i\sin(4\alpha)\cos\beta\sin^2\beta\right].
\end{align}

\end{widetext}

Also the Bir-Pikus Hamiltonian can easily be obtained from the tensor elements above. The matrix elements of the Hamiltonian takes the form
\begin{equation}
    M = \sum_{i,j} M_{ij}^\mathrm{BP} \epsilon_{ij},
\end{equation}
where $\bar\epsilon$ is the strain tensor, and $M_{ij}^\mathrm{BP}$ can be obtained from $M_{ij}$ by the substitution $\{\gamma_1,\gamma_2,\gamma_3\} \to \{-a,\frac{1}{2}b, \frac{1}{2\sqrt 3}d\}$.

\section{Harmonic oscillator Hamiltonian}
\label{app:Level_structure}

The  Hamiltonian we consider has the form
\begin{equation}
    H = \frac{\pi_{x}^2}{2m_-} + \frac{\pi_{y}^2}{2m_+} + \frac{m_-}{2}\omega_x^2 x^2+\frac{m_+}{2}\omega_y^2 y^2,
\end{equation}
with ${\bs \pi} = {\bs p} + e\bs A(\bs r)$, where $\bs A(\bs r) = B_z(-y/2,x/2,0)$, and ${\bs p}=-i\hbar\partial_{\bs r}$ the kinetic momentum.
We insert this expression for ${\bs p}$ and rewrite the Hamiltonian as
\begin{equation}
    H = \frac{ \bar p_{x}^2}{2\mu} + \frac{ \bar p_{y}^2}{2\mu} + \omega_c \bar p_{y} \bar x - \omega_c\bar  p_{x} \bar y +  \frac{\mu}{2}\omega_1^2 \bar x^2+\frac{\mu}{2}\omega_2^2 \bar y^2,
\end{equation}
using the frequencies $\omega_1 = \sqrt{\omega_x^2+\omega_c^2}$, $\omega_2 = \sqrt{\smash[b]{\omega_y^2}+\omega_c^2}$, and $\omega_c = eB_z/2\mu$.
We further rescaled $\bar p_{x} = p_x \sqrt{\mu/m_-}$, $\bar p_{y} = p_y \sqrt{\mu/m_+}$, $\bar x = x\sqrt{m_-/\mu}$, and ${\bar{y}} = y\sqrt{m_+/\mu}$, with $\mu=\sqrt{m_+m_-}$ being the geometric average of the two effective masses.
In this way we rewrote the Hamiltonian as that for an electron with an isotropic mass $\mu$ in an elliptic harmonic potential in the presence of an out-of-plane magnetic field.

There are many ways to diagonalize such a Hamiltonian; we will follow the method outlined in \cite{Qiong_Gui2002}, which leads straightforwardly to
\begin{equation}
    H = \hbar\omega_+\left(a_+^\dagger a_+ + \tfrac{1}{2}\right) + \hbar\omega_-\left(a_-^\dagger a_- + \tfrac{1}{2}\right),
\end{equation}
with $\omega_\pm$ as defined in the main text and
\begin{equation}
    a_\pm = {\bs u}^\pm \cdot \{ \bar x, \bar p_x, \bar y, \bar p_y \},\label{eq:as}
\end{equation}
which obey bosonic commutation relations.
The vectors ${\bs u}^\pm$ read as
\begin{align}
    {\bs u}^\pm = \frac{1}{C_\pm} \{ {} & {} { -i\mu \omega_\pm(\omega_\pm^2 - \omega_y^2-2\omega_c^2)}, \omega_\pm^2-\omega_y^2, \nonumber\\
    {} & {} \quad {-\mu \omega_c(\omega_\pm^2+\omega_y^2)},-2i\omega_c\omega_\pm\},
\end{align}
with
\begin{equation}
    C_\pm = \sqrt{2\hbar \mu\omega_\pm[(\omega_\pm^2-\omega_y^2)^2+4\omega_c^2\omega_y^2]}.\label{eq:cs}
\end{equation}
We can then solve Eq.~(\ref{eq:as}) to express the coordinate and kinetic momentum operators $\{ \bar x, \bar p_x, \bar y, \bar p_y \}$ in terms of the bosonic operators $a_\pm$ and $a_\pm^{\dagger}$,
\begin{align}
    \bar p_x = {} & {} \frac{u^-_3 a_+ - u^+_3 a_-}{2(u^-_3u^+_2-u^-_2u^+_3)} + {\rm H.c.}, \\
    \bar p_y = {} & {} \frac{-u^-_1 a_+ + u^+_1 a_-}{2(u^-_4u^+_1-u^-_1u^+_4)} + {\rm H.c.}, \\
    \bar x = {} & {} \frac{u^-_4 a_+ - u^+_4 a_-}{2(u^-_4u^+_1-u^-_1u^+_4)} + {\rm H.c.}, \\
    \bar y = {} & {} \frac{-u^-_2 a_+ + u^+_2 a_-}{2(u^-_3u^+_2-u^-_2u^+_3)} + {\rm H.c.}.
\end{align}
After scaling back to the original operators $\{ x, p_x, y, p_y \}$, the canonical momenta
\begin{align}
    \pi_x = {} & {} \sqrt{\frac{m_-}{\mu}}\bar p_x - \frac{eB_z}{2} \sqrt{\frac{\mu}{m_+}} \bar y, \\
    \pi_y = {} & {} \sqrt{\frac{m_+}{\mu}}\bar p_y + \frac{eB_z}{2} \sqrt{\frac{\mu}{m_-}} \bar x,
\end{align}
are expressed in terms of the bosonic operators.
Such a form of the momentum operators is very convenient to use in perturbation theory:
In this bosonic framework one can straightforwardly work exclusively in the bosonic Fock space, where no explicit knowledge of the electronic wave functions is required.


%

\end{document}